# The Social Maintenance of Cooperation through Hypocrisy

Todd J Bodnar, Marcel Salathé

*Department of Biology, Penn State University University Park, PA 18062, USA*

***Corresponding authors:**

Marcel Salathé, salathe@psu.edu

W-251 Millennium Science Complex, University Park, PA 18062, USA

Tel: (814) 867-4431

**Cooperation is widespread in human societies, but its maintenance at the group level remains puzzling if individuals benefit from not cooperating. Explanations of the maintenance of cooperation generally assume that cooperative and non-cooperative behavior in others can be assessed and copied accurately. However, humans have a well known capacity to deceive and thus to manipulate how others assess their behavior. Here, we show that hypocrisy - claiming to be acting cooperatively while acting selfishly - can maintain social cooperation because it prevents the spread of selfish behavior. We demonstrate this effect both theoretically and experimentally. Hypocrisy allows the cooperative strategy to spread by taking credit for the success of the non-cooperative strategy.**

The maintenance of cooperation is perhaps one of the greatest enigmas in all of biology. In systems where non-cooperation is more beneficial to an individual than cooperation, why does cooperation persist? Among numerous theories that have been proposed in the past few decades, reciprocity (both direct and indirect)(1, 2), kin selection(3) and group selection(4), and spatial dynamics(5, 6) are the ones most frequently employed to account for this puzzling phenomenon.(7) In addition, punishment and reward have been identified as important contributors to the maintenance of cooperation in human societies.(8-13) The explanatory power for the maintenance of cooperation is usually assessed in game-theoretic models that assume faithful transmission (barring random mutation) of the cooperative or selfish behavior. If the behavior is transmitted genetically, faithful transmission is a valid assumption. However, if the behavior is adopted by imitation, as is generally the case in human affairs, then faithful transmission requires that the behavior can be accurately inferred.(11, 14) Unless the behavior can be directly observed, it must be inferred through other means such as communications, thus allowing for the possibility of deception.

In the context of the game theoretic model of cooperation without direct observation, an individual may act selfishly while claiming to be cooperative. Such hypocritical behavior is indeed observed commonly in psychological studies, where it is argued that hypocrites benefit not only from reaping the rewards of selfish behavior, but also "garner the social and self-rewards of being seen and seeing oneself as upstanding and moral"(15). Here, we argue that hypocrisy can promote the maintenance of cooperation in the population: even if a hypocrite acts selfishly, hypocrisy masks the selfish behavior, preventing it from spreading. We demonstrate this effect with a mathematical model, a computational agent-based simulation model, and an online experiment carried out on Amazon Mechanical Turk(16, 17).

To assess the effect of hypocrisy on the maintenance of cooperation, we use a common model of cooperation, the public goods game(1, 11, 16, 18, 19). In the standard public goods game, players have the option of cooperating by donating to a pool. The donations are then multiplied by a constant and evenly divided between the other players regardless of whether they cooperated or not. In this game, non-cooperating individuals will on average have a higher payoff than cooperating individuals because they don't pay the cost of donating. If strategies with higher payoffs are more often replicated, then the cooperative strategy will become less common over time despite the fact that the population would be best off if everyone would cooperate.

We introduce a modification that concerns the copying process of strategies. It is typically assumed that strategies are faithfully copied (barring mutation), either through biological

transmission(20) (e.g. genetic inheritance), or through cultural transmission(14, 16, 21) (e.g. observational learning). This faithful copying process rests on the assumption that the information transmitted from sender (individual to be copied) to receiver (copying individual) is accurate. The modification that we introduce here is that the information does not need to be accurate. Specifically, we allow for selfish individuals to deceive others about their previous behavior during the imitation process, i.e. to falsely claim that they have been cooperative. We can thus describe an individual by two attributes with respect to cooperation and transmission. First, an individual can be described by the strategy played (cooperative or selfish). Second, and individual can also be described by the strategy transmitted. Standard models of cooperation don't need to make this distinction because the assumed faithful copying process described above - however, in our model, the uncoupling of the strategy played from the strategy transmitted is essential. In order to study the effect of hypocrisy, we assume that there are two types of individuals, non-hypocritical and hypocritical. A non-hypocritical individual will always transmit the strategy that it has just played. A hypocritical individual will always transmit the cooperative strategy, regardless of the strategy that it has just played (Figure 1). We will later modify this definition and assume that a hypocritical individual will always transmit the opposite of the strategy that it has just played. While this latter scenario is perhaps not in line with the common definition of the word hypocrisy, it will provide a conservative baseline for the maintenance of cooperation because it forces hypocritical individuals to transmit the non-cooperative strategy even if it has just played the cooperative strategy (see supplementary material).

We develop a mathematical model (see supplemental 2 for details) that calculates the equilibrium frequency of cooperation c given a constant frequency of hypocrisy h in the population. It can be shown that for a population of any size, c will go to an equilibrium defined by:

$$c = \frac{h}{2 - h} \quad (1)$$

Low levels of random strategy exploration does not fundamentally modify the outcome. These results were confirmed by computational agent-based simulations. A detailed description of the mathematical and computational model is given in the supplementary material.

We tested these predictions experimentally by setting up an online experiment on Amazon Mechanical Turk (AMT), a widely used online platform(16, 17) to recruit participants from a diverse pool. We implemented a public goods game and recruited 414 adults on AMT to participate. Initially, each participant was assigned a random strategy (cooperative or noncooperative), but after each round, participants were allowed to switch strategies. Each round, the participants were shown their payoff in that round, and the payoff and strategy of another player chosen at random. Importantly, participants were not informed that they were playing a public goods game, and the strategies shown were simply labeled A and B. Thus, participants had no prior knowledge of the underlying game design, but were only informed at the beginning that they would be paid according to their performance.

In order to test the effect of hypocrisy on cooperation, we compared 4 different levels of hypocrisy h = 0%, 25%, 50% or 75%, and ran 10 independent trials per hypocrisy level, where each trial is a game with maximally 12 participants playing 10 rounds (participants were prohibited from playing more than one game). If a participant played the noncooperative strategy, her strategy was falsely shown to others as cooperative with probability h. This allowed us to implement hypocrisy without letting participants know that we did, preventing any bias that such knowledge might have introduced. Because the experimental setup of each trial is identical with the exception of h, any difference in the observed frequency of cooperation can be attributed to the effect of hypocrisy.

The results of the online experiment are in agreement with the predictions made by the mathematical and computational models (Figure 2). After random initiation, the cooperation frequency follows the predicted trajectories, attaining near-equilibrium value given by equation (1) after 10 rounds. Using Tukey's honest significance test, we find that the frequency of cooperation is significantly different among all tested levels of hypocrisy (see supplemental table 1).

We've shown here that hypocrisy allows for the maintenance of cooperation in a public goods game. In the standard public goods game without hypocrisy, cooperation cannot be maintained because non-cooperative individuals fare better than cooperative individuals, and their non-cooperative strategy is adopted by others wanting to maximize their payoffs. It's important to note that the addition of hypocrisy to the public goods game does not affect any of these underlying assumptions: selfish individuals are still doing better, and individuals are still adopting the strategies of those around them with the highest payoffs. The only difference is that hypocrisy makes non-cooperative individuals with high payoffs appear cooperative, leading to the adoption of cooperation: Hypocrisy allows the cooperative strategy to spread by taking credit for the success of the non-cooperative strategy.

Punishment of non-cooperative behavior is a well documented solution to problem of cooperation in human communities(8-13, 22). We may hypothesize that hypocrisy has evolved as a means to avoid punishment, providing obvious benefits in addition to those gained by social or self-rewards(15, 23). Our model makes no claim as to why hypocrisy exists, and at what frequencies - it is only concerned with its effect on the maintenance of cooperation. Since hypocrisy is a form of deception, we would expect evolutionary pressure to detect hypocrisy and consequent counter-pressure to avoid detection, leading to an ongoing arm's race. Low levels of hypocrisy detection abilities however do not seem to change our results (see supplementary material).

Supplementary material is available at www.salathegroup.com

**Acknowledgements**

MS acknowledges support from Society in Science - Branco Weiss Fellowship.

**Author Contributions**

TJB and MS conceived the project, designed the experiment, analyzed the data, and wrote the paper. TJB developed the models and implemented the AMT experiments.

**Competing Interests**

The authors declare no competing financial interests.

**Human Subject Experiments**

The AMT experiments were approved by the Penn State Institutional Review Board on October 1, 2012 (Protocol ID 41024).

# FIGURE CAPTIONS

**Figure 1**

A visual representation of the reported behavior (background) and the actual behavior (circle) for each combination of game and reporting strategies. Honest individuals report the same strategy that they actually did while hypocrites always report that they cooperated. (Key: Red = Defect, Blue = Cooperate)

**Figure 2**

Dynamics predicted by the mathematical (solid lines) and agent based (dashed lines) models compared to the observed frequency of cooperation observed in the AMT experiments (circles). Transparent circles are the frequencies from each of the experimental groups, while the solid circles are the means of all of the experiments per level of hypocrisy and round. Horizontal jittering added for better visibility.

**FIGURES**

**Figure 1**

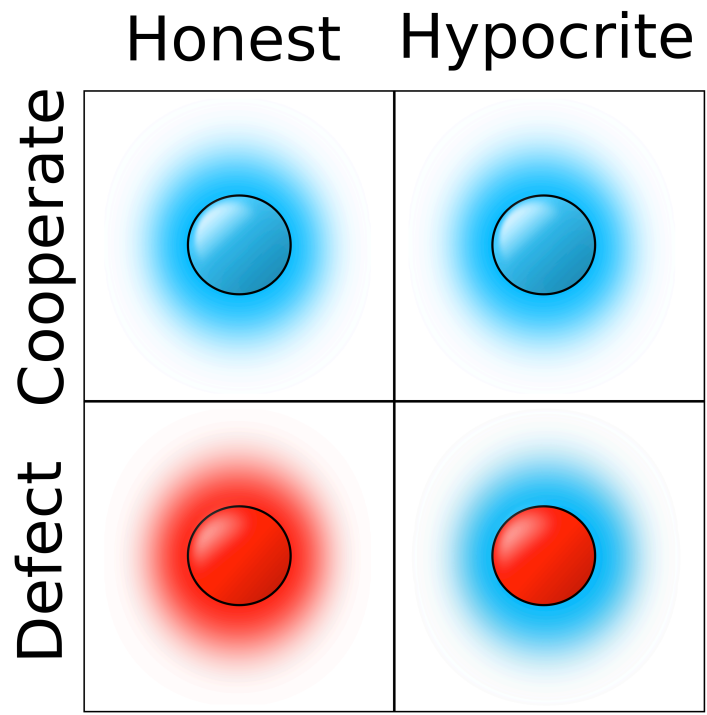

**Figure 2**

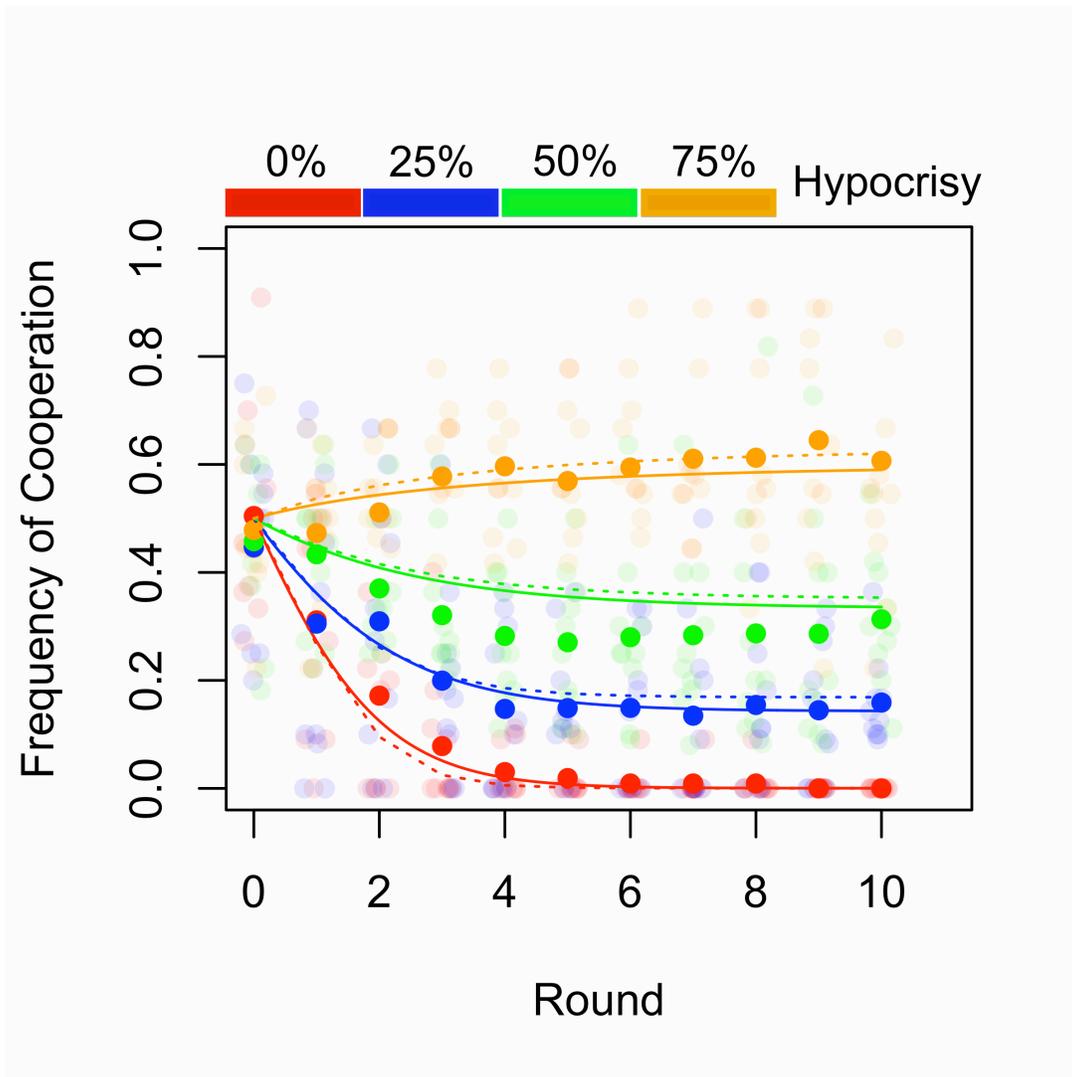